\documentclass[12pt,preprint]{aastex}

\shorttitle{SiO masers around AH Sco} \shortauthors{Xi Chen et
al.}

\begin{document}


\title {VLBI Observations of SiO Masers around AH Scorpii }
\author
 {Xi ~Chen\altaffilmark{1} \& Zhi-Qiang ~Shen\altaffilmark{1}}

\altaffiltext{1} {Shanghai Astronomical Observatory, 80 Nandan
Road, Shanghai 200030, P.R.China}

\label{firstpage}


\begin{abstract}

We report the first Very Long Baseline Array (VLBA) observations
of 43 GHz $v$=1, $J$=1--0 SiO masers in the circumstellar envelope
of the M-type semi-regular supergiant variable star AH Sco at 2
epochs separated by 12 days in March 2004. These high-resolution
VLBA images reveal that the distribution of SiO masers is roughly
on a persistent elliptical ring with the lengths of the major and
minor axes of about 18.5 and 15.8 mas, respectively, along a
position angle of $150^{\circ}$. And the red-shifted masers are
found to be slightly closer to the central star than the
blue-shifted masers. The line-of-sight velocity structure of the
SiO masers shows that with respect to the systemic velocity of
$-6.8$ km s$^{-1}$ the higher velocity features are closer to the
star, which can be well explained by the simple outflow or infall
without rotation kinematics of SiO masers around AH Sco. Study of
proper motions of 59 matched features between two epochs clearly
indicates that the SiO maser shell around AH Sco was undergoing an
overall contraction to the star at a velocity of $\approx$13 km
s$^{-1}$ at a distance of 2.26 kpc to AH Sco. Our 3-dimensional
maser kinematics model further suggests that such an inward motion
is very likely due to the gravitation of the central star. The
distance to AH Sco of 2.26$\pm$0.19 kpc obtained from the
3-dimensional kinematics model fitting is consistent with its
kinematic distance of 2.0 kpc.

\end{abstract}


\section{Introduction}

AH Scorpii (AH Sco) is a semi-regular variable with an optical
period of 714 days (Kukarkin et al. 1969) and a spectral type of
M5Ia-Iab (Humphreys 1974). The systemic velocity of AH Sco is
estimated to be about $-7$ and $-3$ km s$^{-1}$ based on the
observations of OH maser (Baudry, Le Squeren \& L\'{e}pine 1977)
and H$_{2}$O maser (L\'{e}pine, Pase de Barros \& Gammon 1976),
respectively. The distance to AH Sco remains uncertain. A
photometric distance of 4.6 kpc has been derived by Humphreys \&
Ney (1974) based on their infrared data. However, Baudry, Le
Squeren \& L\'{e}pine (1977) led to a photometric distance of 2.6
kpc under an assumption of an absolute visual magnitude of $-5.8$
for Iab stars, whereas its kinematic distance was about
$1.5\sim2.0$ kpc for its systemic velocity $-5.5\sim-7.5$ km
s$^{-1}$.

Late type stars often exhibit circumstellar maser emission in
molecules OH, H$_{2}$O, and SiO. The supergiant variable AH Sco is
such a star that has been detected strong maser emission with
single-dish in all three species (e.g. L\'{e}pine, Pase de Barros
\& Gammon 1976; Baudry, Le Squeren \& L\'{e}pine 1977; Balister et
al. 1977, G\'{o}mez Balboa \& L\'{e}pine 1986). The
interferometric observations of these masers would be useful in
determining the structure and kinematics of the circumstellar
envelop (CSE) and understanding the physical circumstance and mass
loss procedure for this supergiant variable. Unfortunately, there
has been so far no any published interferometric map of OH,
H$_{2}$O and SiO masers toward this source.

Especially, SiO masers provide a good probe of the morphology of
CSE and kinematics of gas in the extended atmosphere which is a
complex region located between the photosphere and the inner dust
formation shell. Previous VLBI experiments have demonstrated
ringlike configurations (e.g. Diamond et al. 1994; Greenhill et
al. 1995; Boboltz, Diamond \& Kemball 1997; Yi et al. 2005, Chen
et al. 2006), or elliptical distributions (e.g. Boboltz \& Marvel
2000; S\'{a}nchez et al. 2002; Boboltz \& Diamond 2005) of SiO
masers, and also revealed complex kinematics in SiO maser regions,
e.g. contraction and expansion at the different phase of stellar
pulsation (Boboltz, Diamond \& Kemball 1997; Diamond \& Kemball
2003; Chen et al. 2006) and even rotation (Boboltz \& Marvel 2000;
Hollis et al. 2001; S\'{a}nchez et al. 2002; Cotton et al. 2004;
Boboltz \& Diamond 2005). The ringlike or elliptical distribution
that is assumed to be centered at the stellar position with a
radius of 2--4 R$_{\ast}$ suggests that SiO masers are amplified
in tangential rather than radial path.

In this paper, we present the first VLBI maps of SiO maser
emission toward AH Sco observed at two epochs separated by 12 days
in March 2004. The observations and data reduction are described
in $\S$ 2; results and discussions are presented in $\S$ 3,
followed by conclusions in $\S$ 4.

\section{Observations and data reduction}

The observations of the $v=$1, $J$=1--0 SiO maser emission toward
AH Sco ($\alpha=$17$^h$11$^m$16.98$^s$,
$\delta=-32\arcdeg19\arcmin31.2\arcsec$, J2000) were performed at
two epochs on March 8, 2004 (hereafter epoch A) and March 20, 2004
(hereafter epoch B) using the 10 stations of the Very Long
Baseline Array (VLBA) of the NRAO\footnote[1]{The National Radio
Astronomy Observatory is a facility of the National Science
Foundation operated under cooperative agreement by Associated
Universities, Inc.}. A reference frequency of 43.122027 GHz was
adopted for the $v$=1, $J$=1--0 SiO transition. The data were
recorded in left circular polarization in an 8 MHz band and
correlated with the FX correlator in Socorro, New Mexico. The
correlator output data had 256 spectral channels, corresponding to
a velocity resolution of 0.22 km s$^{-1}$. The system temperatures
and sensitivities were on the order of 150 K and 11 Jy K$^{-1}$,
respectively, for both epochs.

For the data reduction, we followed the standard procedure for
VLBA spectral line observations using the Astronomical Image
Processing System (AIPS) package. The bandpass response was
determined from scans on the continuum calibrator (NRAO530). The
amplitude calibration was achieved using the total-power spectra
of AH Sco based on the ``template spectrum'' method. The template
spectra at each epoch were obtained from the Mauna Kea (MK)
station at a high elevation. A zenith opacity of about 0.05 at MK
station estimated from the variation of system temperature with
zenith angle was applied to correct the atmosphere absorption for
each epoch. Residual group delays determined from a fringe fitting
to the continuum calibrator were applied to the spectral line
data. Residual fringe-rates were obtained by performing a
fringe-fitting on a reference channel (at V$_{\rm LSR}=-$9.4 km
s$^{-1}$), which has a relatively simple structure in the maser
emission. An iterative self-calibration on the reference channel
was performed to remove any structural phase. The solutions of
fringe-fitting and self-calibration were then applied to the whole
spectral line data. Image cubes were produced for all the velocity
channels between 14 and $-$18 km s$^{-1}$ with a synthesized beam
of 0.69 mas $\times$ 0.20 mas at a position angle $-6^\circ$.
Off-source rms noise ($\sigma_{\rm rms}$) in channel maps ranges
from 20 mJy beam$^{-1}$ in the maps with weak or no maser emission
to 50 mJy beam$^{-1}$ in the maps containing strong maser
emission. The flux densities, and positions in right ascension
(R.A.) and declination (Dec.) for all emission components with
intensity above 8 $\sigma_{\rm rms}$ in each channel maps were
determined by fitting a two-dimension Gaussian brightness
distribution using the AIPS task SAD. Errors in R.A. and Dec.
obtained from above fitting procedure range from 1 $\mu$as for
components with high SNR, to 168 $\mu$as for components with low
SNR, and the typical uncertainty of the fitted position of maser
components was smaller than 10 $\mu$as.

The remaining analysis of the maser component identifications was
performed outside of the AIPS package. As described in our
previous work (Chen et al. 2006), a maser spot is a single
velocity component of the maser emission in each velocity channel
map; a maser feature is a group of the maser spots within a small
region in both space and Doppler velocity, typically 1 AU and 1 km
s$^{-1}$, and is expected to be a physical feature consisting of a
single gas clump. In order to study the characteristics of SiO
masers, it is necessary to identify maser features for each epoch.
The maser spots in different channels were deemed as the same
feature according to the criterion that these spots appear in at
least three adjacent channels and lie within an angular separation
of 0.5 mas. Finally, 82 and 87 maser features were identified for
epochs A and B, respectively.

\section{Results and Discussions}

\subsection{The spatial structure of the SiO masers}

The full lists of parameters for each identified feature are given
in Tables 1 and 2, for epochs A and B, respectively. We fit a
Gaussian curve to the velocity profile of a feature containing at
least four spots to determine V$_{\rm LSR}$ at the peak of
velocity profile. For some features which can not be well
represented by a Gaussian profile (labelled by a ``$\ast$'' in
Tables 1 and 2), intensity weighted mean V$_{\rm LSR}$ was
adopted. The velocity range across the feature $\Delta u$, is
defined to be the difference between the maximal and minimal
velocities of the spots in the feature. Feature positions ($x$,
$y$) in R.A. and Dec. were determined from an intensity weighted
average over maser spots in the feature. The uncertainty
($\sigma_{x}, \sigma_{ y}$) of a feature position was defined as
squared root of the square sum of (1) the mean spot distance from
the defined feature position and (2) the mean measurement error of
the spot positions. The weights proportional to the intensity of
the spot were applied in the uncertainty estimation. The typical
position uncertainties of features are 0.01 mas and 0.02 mas for
R.A. and Dec., respectively.  The positions are measured with
respect to the reference feature at (0, 0) for aligning the maps
in the two epochs (labelled by ``R'' in Tables 1 and 2; see Sect.
3.2.2). The distance of a maser feature, $r$, is measured with
respect to the fitted center obtained from the ellipse model
fitting to the distribution of maser features (see below). The
flux density of the brightest spot in each feature was deemed as
its peak flux density, P.

In two top panels of Figure 1, we compare the total power imaged
by the VLBA (open circle) to the total power (solid line) obtained
from the MK antenna for each epoch. The total power imaged by VLBA
is obtained by summing all fitted flux of spots belonging to
features. The fractional power representing the ratio of the total
flux imaged by VLBA to the total flux of maser emission is shown
in two bottom panels. The fraction is mostly between 0.4 and 0.8.
That is, on average about $\sim$60\% of the total luminosity of
masers was detected in our observations. Actually, such a
fractional power reflects the degree of extension of the maser
emission. If the apparent sizes of maser components are larger
than that of the interferometric beam, the maser would be partly
resolved (i.e. the fractional power is less than 1) and the
fractional power should decrease with the increase of apparent
size of masers. The typical size of maser spots estimated from
geometric mean of sizes of the major and minor axes of the spots,
which were obtained by fitting to an elliptical Gaussian
brightness distribution in the CLEAN map, is 0.5 mas. This typical
scale size is slightly larger than the geometric mean of the VLBA
beam of 0.4 mas. Thus $\sim$40\% missing flux in our map is mainly
due to the high spatial resolution of the interferometric array.

Figure 2 shows the distributions of maser features toward AH Sco
for the two epochs. These high resolution VLBI images reveal a
persistent elliptical structure of SiO masers around AH Sco druing
an interval of 12 days. We characterized this morphology by
performing a least-squares fit of an ellipse to the distribution
of masers weighted by the flux density of each feature for each of
two epochs. The best-fitting ellipses and the ellipse centers are
also shown in Figure 2. The lengths of the major and minor axes
were found to be 18.6 and 15.7 mas for epoch A, and 18.4 and 15.9
mas for epoch B, respectively, with the major axis of the ellipse
oriented similarly at $\sim150^{\circ}$ at both epochs. And the
fitted centers of the elliptical distributions are almost the same
at both epochs of $-7.9$ and 5.8 mas in R.A. and Dec.,
respectively. At a distance to AH Sco of 2.26 kpc (see Sect.
3.2.3), the distribution of SiO masers corresponds to about
$42\times35$ AU for both two epochs. However, an axial ratio of
1.18 suggests that the distribution of maser features around AH
Sco can be viewed approximately as the ringlike structure with an
average diameter of about 17.2 mas (obtained from the geometric
average of the major and minor axes for both epochs).

The simultaneous near-infrared interferometry and radio
interferometry imaging of circumstellar SiO maser for late type
stars have been done recently (e.g. Boboltz \& Wittkowski 2005;
Wittkowski et al. 2007; Cotton et al. 2004, 2006). These
observations reveal that the ratio of the maser ring radius to the
photospheric radius of the central star is about $1.5-4.0$.
Unfortunately, there has been no any published photospheric radius
measurement for AH Sco. Thus we can not directly compare SiO maser
ring radius with stellar radius for AH Sco.

We also notice that the red-shifted SiO masers lie slightly closer
to the center than the blue-shifted masers (see Figure 2). This
can be seen more clearly in Figure 3, showing the maser feature
distance from the fitted center versus its line-of-sight (LOS)
velocity (see Sect. 3.2.1). This phenomenon seems to be explained
under the assumption that maser gas was undergoing infall to the
central star. This is because along the same LOS path the red- and
blue-shifted masers would appear in front of and behind the star,
respectively, as long as the coherence path lengths satisfy the
requirement of maser excitation, under the condition that the
maser gas was undergoing infall to the central star during our
observations. However the blue-shifted masers generated behind
star would be obscured by the stellar disc projected on the LOS,
while the red-shifted masers would not. Thus only red-shifted
maser emission will be seen closer to the center. Actually, we
have confirmed that the SiO maser shell contracts to the star
during our observations in Sect. 3.2.2. Moreover, from Figure 3,
some systemic masers with velocity $-7$ km s$^{-1}$ (see Sect.
3.2.1) locate at the distance of 7 mas which can be viewed as an
upper limit to the photospheric radius, then the extreme
blue-shifted masers with the distance of less than 7 mas would be
obscured. This is consistent with our data (see Figure 3) showing
that all blue-shifted masers locate at the distance of larger than
7 mas.

\subsection{The kinematics of the SiO masers}

\subsubsection{The kinematics obtained from LOS velocities}

From Figure 2, we notice that there appears a velocity
gradient at both epochs, with the bluest- and reddest-shifted
maser features lying closer to the center of the distribution than
those with intermediate velocities. To verify this, we plotted in
Figure 3 feature distance from the fitted center (marked by the
red star in Figure 2) versus its LOS velocity for both epochs.
Both epochs appear to have the same distribution with a peak near
the velocity of $-7$ km s$^{-1}$ and decreasing maser distance
with increasing deviation of velocity from this peak velocity.
This has also been seen in some OH maser sources (e.g. Reid et al.
1977, Chapman \& Cohen 1986), H$_{2}$O maser sources (e.g. Yates
\& Cohen 1994) and SiO maser sources (e.g. Boboltz \& Marvel 2000;
Wittkowski et al. 2007). A widely used simple model to explain
this phenomenon is that of a uniformly expanding thin shell (e.g.
Reid et al. 1977; Yates \& Cohen 1994; Wittkowski et al. 2007). In
this model the projected distance $r$ (in the fourth column of
Tables 1 and 2) of a maser on the shell from the center is related
to its LOS velocity $V_{\rm LSR}$ by the expression

\begin{equation}\label{13}
  (\frac{r}{r_{s}})^2+(\frac{V_{\rm LSR}-V_{*}}{V_{exp}})^2=1,
\end{equation}
where $V_{*}$ is the systemic velocity of maser source, $r_{s}$ is
the shell radius and V$_{exp}$ is the expanding velocity.
Apparently, this model traces an ellipse on the $r-V_{\rm LSR}$
plot.

The uniformly expanding thin shell model was used to characterize
the expansion or contraction kinematics of a circular maser
distribution. For the case of AH Sco, even though we characterize
the maser distribution as an ellipse, an axial ratio of $\sim$1.2
suggests that the distribution of maser features is approximately
a circular structure as discussed in Section 3.1. Thus, we can
also apply the uniformly expanding thin shell model to AH Sco.
Moreover, most maser features locate in the northwest and
southeast (i.e. the direction of major axis), and only few maser
features locate in the northeast and southwest (i.e. the direction
of minor axis). These make the assumption of uniformly
expansion/contraction in all directions of the uniformly expanding
thin shell model to be suitable for the case of AH Sco. We
performed a least-squares fit of the uniformly expanding thin
shell model to the distribution of epochs A and B. The LSR stellar
velocity of AH Sco, which is estimated to be about $-7$ and $-3$
km s$^{-1}$ based on the observations of OH maser (Baudry, Le
Squeren \& L\'{e}pine 1977) and H$_{2}$O maser (L\'{e}pine, Pase
de Barros \& Gammon 1976), has not been measured particularly well
yet. Thus the LSR stellar velocity $V_{*}$, together with the
shell radius $r$ and expansion velocity V$_{exp}$, is treated as a
free parameter in the fitting procedure. The values of $V_{*}$,
V$_{exp}$ and $r_s$ were found to be $-6.8\pm$0.5 km s$^{-1}$,
18.8$\pm$2.0 km s$^{-1}$ and 9.3$\pm$0.1 mas, respectively, for
both epochs, where the uncertainties are their standard errors.
The best fitted $r-V_{\rm LSR}$ ellipse is also plotted in Figure
3. The systemic velocity of AH Sco of $-6.8$ km s$^{-1}$ is
consistent with the value of $-7$ km s$^{-1}$ from the OH maser
observations. The shell radius of 9.3 mas determined from above
model fitting is larger than the radius of maser distribution of
8.6 mas (see Sect. 3.1). A note is that the definition of the
shell radius determined from the uniformly expanding thin shell
model is different from that of the radius of the maser
distribution. The shell radius reflects the scale of a
3-dimensional maser spherical shell, whereas the radius of maser
distribution reflects the scale of maser distribution in the sky
plane, and is the projected size on the sky plane of the
3-dimensional spherical shell. Thus it is not surprising that the
shell radius is a bit larger than the radius of maser
distribution.

For comparison, the escape velocity calculated at the shell
radius of 9.3 mas, assuming a typical mass of 10 $M_{\odot}$ for
supergiant and a distance to AH Sco of 2.26 kpc, is about 29 km
s$^{-1}$. Thus, the expansion/contraction velocity at the
location of SiO maser shell is less than the corresponding escape
velocity, suggesting that the maser gas is still gravitational
bound to the star. However, the current fitting can not tell the
sign of V$_{exp}$ term (as can be seen in Eq. (1)), and thus can
not differentiate between expansion and contraction of the maser
shell. The dominant expansion or contraction of maser shell can be
clarified by the SiO maser proper motion analysis to be discussed
below.

\subsubsection{Maser Proper Motions}

We can study proper motion and the kinematics of the CSE of AH Sco
by tracing the matched features that appeared in both epochs.
Because the absolute position of the phase center in each image is
not kept during the data reduction, we must align two-epoch maps
for studying the proper motion. The feature used for registration
is the one with a velocity V$_{\rm LSR}\approx-9.3$ km s$^{-1}$
(labelled by ``R'' in Tables 1 and 2) at both epochs. And then we
shift the coordinate frames for both epochs to align the origin
(0, 0) with this feature. At an assumed distance of 2.6 kpc
(Baudry et al. 1977) and a maximum expansion/contraction velocity
of 20 km s$^{-1}$ (see Sect. 3.2.1), the maser proper motion
should be less than 0.05 mas in an interval of 12 days. Thus we
can match these features from one epoch to another epoch using the
criterion that the angular separation of the matched features
between two epochs should not exceed 0.15 mas after allowing the
maximum position uncertainty of (0.05, 0.10) mas (see Tables 1 and
2) and they have similar velocity profile and flux density. As a
result, we identified 59 commonly matched maser features between
two epochs (see Tables 1 and 2). Actually, using a reference
feature located on the maser shell to align maps for two epochs
could introduce a constant offset vector representing the motion
of the reference feature in the individual maser proper motions.
We assumed that the vector-average of the proper motions for all
the matched features represents the motion of the aligned feature.
In order to present a better representation of the real motions of
individual features, the mean proper motion was subtracted from
each of the determined proper motion vectors. The proper motions
of matched maser features are shown in Figure 4. Here we adopt the
distance to AH Sco of 2.26 kpc (see Sect. 3.2.3) for denoting the
velocity values of the proper motions.

From Figure 4, we can clearly see that the maser shell shows an
overall contraction toward the central star. In order to better
characterize the net contraction of the masers, we computed the
separations between pairwise combinations of features. This
technique has previously been applied to analyse proper motions of
OH masers (Chapman, Cohen \& Saika 1991; Bloemhof, Reid \& Moran
1992), H$_{2}$O masers (Boboltz \& Marvel 2007), and SiO masers
(Boboltz, Diamond \& Kemball 1997; Chen et al. 2006), and has no
dependence on the alignment of maps. The procedure involves
computing the angular separation between two features at the first epoch
and the separation between the corresponding two features at the
second epoch. The difference between the two values of separation
is referred to as the pairwise separation. The procedure is
repeated for all the possible pair combinations. However, the
inclusion of all possible pair combinations often results in
decreasing toward zero due to the bias caused by calculating pairs
of closely spaced maser features. For the sake of clarity, and to
determine representative values for the angular shifts due to the
contraction, we have included only those pairs separated by more
than 9 mas (corresponding to the radius of maser distribution). We
obtained the mean value of these pairwise separations of $-0.039\pm
0.002$ mas, in an interval of 12 days, corresponding to a proper
motion of $-1.186\pm 0.061$ mas yr$^{-1}$ or a velocity of
$-12.7\pm 0.7$ km s$^{-1}$ at a distance of 2.26 kpc, where the
uncertainties are the standard errors. The negative
value of proper motion implies an overall contraction of the maser
shell. The contraction value of $1.186\pm 0.061$ mas yr$^{-1}$ of
maser shell derived from the pairwise separation is significantly
less than the scalar-averaged value of SiO proper motions of
$1.96\pm 0.15$ mas yr$^{-1}$ (where its uncertainty is the
standard error). This is because some of maser proper
motions do not completely point to the center or even a few proper
motions show outflow motion as can be seen in Figure 4.

The contraction of SiO maser shell has been reported in two Mira
variables R Aqr (Boboltz, Diamond \& Kemball 1997) and TX Cam
(Diamond \& Kemball 2003) and one red supergiant VX Sgr (Chen et
al. 2006; 2007). Our observations provide an inward motion of SiO
maser shell around another red supergiant AH Sco. In Table 3 we
list these four sources. We also estimated the stellar optical
phase of AH Sco at our observation sessions to be $\phi\simeq$
0.55 (i.e. at the optical minimum phase) based on the American
Association of Variable Star Observers (AAVSO) data.
Interestingly, the optical phase at which the SiO maser shell
around the red supergiant AH Sco contracts is nearly the same as
that seen in other three sources: VX Sgr ($\phi=0.75-0.80$; Chen
et al. 2006), R Aqr ($\phi=0.78-0.04$; Boboltz, Diamond \& Kemball
1997), TX Cam ($\phi=0.50-0.65$; Diamond \& Kemball 2003). This
infers that the contraction of the SiO maser shell would occur
during an optical stellar phase of $0.5-1$, which agrees with the
previous conclusion reported by Chen et al. (2006) and the
theoretical kinematical model results of Humphreys et al. (2002).
Moreover, from Table 3 we can find that the contraction velocity
of about 13 km s$^{-1}$ of maser shell around AH Sco is the
largest among the four sources.

\subsubsection{3-dimensional kinematics model for SiO masers}

In order to estimate further the kinematical parameters of SiO
masers and the distance to AH Sco, we made model-fitting to
analyze spatial distribution and proper motion of SiO maser
features as done by Gwinn, Moran \& Reid (1992) and  Imai et al.
(2000; 2003). The model fitting requires to minimize the squared
sum of the differences between the observed and model velocities,
\begin{equation}\label{2}
\chi^2=\sum_i\{
   \frac{[\mu_{ix}-V_{ix}/(a_{0}d)]^2}{\sigma_{ix}^2}+\frac{[\mu_{iy}-V_{iy}/(a_{0}d)]^2}{\sigma_{iy}^2}+\frac{[u_{iz}-V_{iz}]^2}{\sigma_{iz}^2}\},
\end{equation}
where, $\mu_{ix}$ and $\mu_{iy}$ are the observed proper motions
in R.A. and Dec., respectively, and $u_{iz}$ the observed velocity
along LOS, $d$ the distance to the maser source, $a_{0}\equiv4.74$
km s$^{-1}$ mas $^{-1}$ yr kpc$^{-1}$, ($\sigma_{ix}, \sigma_{iy},
\sigma_{iz}$) the standard deviations of the observed velocity
vectors which are determined in the similar manner of Imai et al.
(2002). In this work, we assume a spherically expanding flow in
SiO maser region, thus the model velocity vector $\textbf{V}_i$
(including $V_{ix}$, $V_{iy}$ in R.A. and Dec., and $V_{iz}$ in
LOS) for the $i$th maser feature can be expressed as

  \begin{equation}\label{01}
 \mathbf{V}_i=\mathbf{V_{0}}+V_{exp}(i)\frac{\mathbf{r}_{i}}{r_{i}}\mathrm{,}
\end{equation}

\begin{equation}\label{02}
   \mathbf{r}_{i}=\mathbf{x}_{i}-\mathbf{x}_{0} \ (\mathrm{or}\ r_{ix}=x_{i}-x_{0}, r_{iy}=y_{i}-y_{0},
   r_{iz}=z_{i}),
\end{equation}
where $\textbf{V}_0$ ($v_{0x}$, $v_{0y}$, $v_{0z}$) is a systemic
velocity vector of the stellar system, reflecting the motion of
the central star; $V_{exp}(i)$ is an expanding velocity as a
function of the distance from the origin of the flow, r$_{i}$;
\textbf{x}$_0$ ($x_{0}$, $y_{0}$, 0) is the position vector of the
flow origin with respect to the position of reference maser
feature (here, we assumed that the origin of the flow was at the
star, whose positions of $x_{0}=-7.94$ mas and $y_{0}=-5.83$ mas
have been derived from the least-squares fit of an ellipse to the
distribution of masers in Sect. 3.1.); and ($x_{i}$, $y_{i}$) is
an observed position of maser feature on the sky plane; the
position of a maser feature along the LOS, $z_{i}$ is estimated as
one of the free parameters too (Imai et al. 2000). In the model
fitting procedure, we adopted the expanding velocity as
$V_{exp}(i)=V_{1}(r_{i}/r_{0})^\alpha$, where $V_{1}$ is the
expansion velocity at a unit distance of $r_{0}\equiv 10$ mas, and
$\alpha$ a power-law index indicating the apparent acceleration or
deceleration of the flow. Moveover, we excluded those maser
features with large positive expansion velocities in the fitting.
Finally, we used 48 proper motion data and obtained the best
solutions with their standard errors, which are given in Table 4.

From Table 4, we can see that a negative expansion velocity of
$-14.1\pm1.4$ km s$^{-1}$ at a shell radius of 10 mas was
estimated from the best-fit model, supporting the presence of a
real contracting flow in the SiO maser region around AH Sco. This
is consistent with the conclusion derived from the pairwise
separation calculation in Sect. 3.2.2. For comparison, we also
estimate the corresponding velocity of $-15.0\pm1.5$ km s$^{-1}$
at the mean SiO maser radius of 9 mas according to the power-law
index of the acceleration of the flow $\alpha$ obtained in our
model. However, the velocity of $-15.0\pm1.5$ km s$^{-1}$ seems
larger than that of $-12.7\pm0.7$ km s$^{-1}$ obtained from the
pairwise separation analysis, which is because that we excluded
maser features showing outflow motion in the 3-dimensional
kinematics model fitting procedure. And the velocity of
$15.0\pm1.5$ km s$^{-1}$ is slightly smaller than that of
18.8$\pm$2.0 km s$^{-1}$ obtained from a least-squares fit of the
uniformly expanding thin shell model to SiO maser LOS velocity
structure (see Sect. 3.2.1), which is due to that some maser
proper motions used in model fitting still deviate from the
originating point of the inflow (i.e. the position of central
star; see Figure 4), whereas the model fitting method involves one
critical assumption that velocity vectors are in the radial
direction from the commonly originating point of the flow. The
systemic velocity along LOS of $v_{0z}=-5.2$ km s$^{-1}$ from the
3-dimensional kinematics model is also roughly consistent with
that of $-6.8$ km s$^{-1}$ derived from the uniformly expanding
thin shell model in Sect. 3.2.1, suggesting that the assumption of
the origin of the flow located at the star is reasonable. More
interestingly, we obtain a negative power-law index of the
acceleration of the flow, $\alpha=-0.54\pm0.16$, indicating that
the contracting flow was accelerating in the SiO maser region with
a similar form of $v\sim r^{-0.5}$ of the gravitational
contraction. Thus the 3-dimensional maser kinematics model
suggests that the infall motions of SiO masers can be achieved
under the gravitational effect of the central star.

The velocity gradient across the SiO maser region usually used in
the numerical simulation of SiO masers (e.g.  Doel et al. 1995;
Humphreys et al. 2002) can be expressed by
\begin{equation}\label{14}
  \rm \varepsilon=\frac{dlnV}{dlnr}=\frac{rdV}{Vdr}.
\end{equation}
A value of $\varepsilon=0$ corresponds to a constant velocity
expansion, while $\varepsilon\geq1$ corresponds to a velocity
field with large radial accelerations. A value of $\varepsilon=1$
was usually adopted in the SiO maser numerical simulations (e.g.
 Doel et al. 1995). Chapman \& Cohen
(1986) estimated the velocity gradient $\varepsilon$ for the OH,
H$_{2}$O and SiO maser emission around VX Sgr and found that
$\varepsilon\approx0.2$ in the 1612 MHz OH maser region,
$\varepsilon\approx0.5$ in the region of the H$_{2}$O and mainline
OH masers, and $\varepsilon\approx1$ in SiO maser region. However,
the power-law index of the acceleration of the flow
($\alpha=-0.54\pm0.16$) derived from our best-fit kinematic model
suggests a negative velocity gradient value of $\varepsilon=-0.54$
in SiO maser region around AH Sco. This is different from those
flows that apparently exhibit the accelerations in their SiO maser
kinematics (e.g. VX Sgr, Chapman \& Cohen 1986; S Ori, Wittkowski
et al. 2007) and the positive velocity gradient value used in the
maser simulations. The content of maser simulations is beyond this
work. However, we think that such a negative velocity gradient
$\varepsilon$ adpoted in maser simulation may be necessary for
understanding the SiO maser emission, especially during the infall
stage of the SiO maser shell.

The 3-dimensional kinematics model fitting shows a best solution
for the distance to AH Sco of $2.26\pm0.19$ kpc. This distance
value seems reasonable. Firstly, this distance of AH Sco is in a
good agreement with its estimated `near' kinematic distance of
about 2.0 kpc at the systemic velocity of $-6.8$ km s$^{-1}$ with
the adopted galactic constants, R$_{\odot}=$ 8.5 kpc and
$\Theta_{\odot}=$ 220 km s$^{-1}$. Secondly, at the distance of
2.26 kpc, the scalar-averaged value of $1.96\pm0.15$ mas yr$^{-1}$
of proper motions of matched features (shown in Figure 4) would
correspond to a velocity of $21.0\pm1.6$ km s$^{-1}$. This
velocity is consistent with the expansion/contraction velocity of
$18.8\pm2.0$ km s$^{-1}$ in SiO maser region obtained from a
least-squares fit of the uniformly expanding thin shell model to
LOS velocity structure. Thus we adopt the distance to AH Sco of
2.26 kpc throughout this work.

\section{Conclusions}

We summarize the main results obtained from 2-epoch (at an
interval of 12 days) monitoring observations of the 43 GHz $v=1,
J=1-0$ SiO maser emission toward AH Sco performed in March 2004,
corresponding to a stellar optical phase of $\sim$ 0.55.

\begin{enumerate}
\renewcommand{\theenumi}{(\arabic{enumi})}

  \item
  Our observations revealed a persistent elliptical structure of SiO
masers with the sizes of the major and minor axes of about 18.5
and 15.8 mas, respectively, along a position angle of
$150^{\circ}$. We notice that the red-shifted SiO maser
emission lies slightly closer to the center than the blue-shifted
one.
  \item

The LOS velocity structure of the SiO masers shows a velocity
gradient at both epochs, with masers towrad the blue- and
red-shifted ends of the spectrum lying closer to the center of the
maser distribution than masers at intermediate velocities, which
can be explained by the outflow or infall kinematics of SiO maser
shell. By analyzing the uniformly expanding thin shell model to
the LOS velocity  of SiO masers, we estimated the
expansion/contraction velocity of about 19 km s$^{-1}$ in SiO
maser region around AH Sco.

  \item
The proper motions of 59 matched features between two epochs show
that the SiO maser shell around AH Sco was undergoing inward
motion to the central star. Computing pairwise separation of these
matched features, we obtained that the maser shell contracts
toward AH Sco with a velocity of about $13$ km s$^{-1}$ at a
distance to AH Sco of 2.26 kpc.  The stellar optical phase of red
supergiant AH Sco is very close to that of Mira variables R Aqr
and TX Cam and red supergiant VX Sgr when the SiO maser shell
contracts. And the contraction velocity of about 13 km s$^{-1}$ of
maser shell around AH Sco is the largest one among the known four
sources showing contraction of SiO maser shell.

 \item

We made a 3-dimensional kinematics model to analyze spatial
distribution and proper motion of SiO maser features. The
3-dimensional maser kinematics model further suggested that the
contraction of SiO maser shell around AH Sco is mainly due to the
gravitation of the central star. And the distance to AH Sco of
$2.26\pm0.19$ kpc estimated from this kinematics model fitting is
consistent with the kinematic distance of 2.0 kpc at the systemic
velocity of AH Sco of $-7$ km s$^{-1}$.

\end{enumerate}

\acknowledgements

We thank an anonymous referee for helpful comments that improved
the manuscript. We also acknowledge with thanks data from the
AAVSO International Database based on observations submitted to
the AAVSO by variable star observers worldwide.

This work was supported in part by the National Natural Science
Foundation of China (grants 10573029, 10625314, and 10633010) and
the Knowledge Innovation Program of the Chinese Academy of
Sciences (Grant No. KJCX2-YW-T03), and sponsored by the Program of
Shanghai Subject Chief Scientist (06XD14024) and the National Key
Basic Research Development Program of China (No. 2007CB815405).

X. Chen thanks the support by the Knowledge Innovation Program of
the Chinese Academy of Sciences. ZQS acknowledges the support by
the One-Hundred-Talent Program of Chinese Academy of Sciences.

\clearpage

\begin{deluxetable}{lrrrrrrrrrr}

\tabletypesize{\tiny} \setlength{\tabcolsep}{0.11in}
\tablewidth{0pt} \tablecaption{\footnotesize 43 GHz SiO maser
features around AH Sco observed by VLBA on March 8, 2004.}

\tablehead{ID&\multicolumn{1}{r}{V$_{LSR}$}&$\Delta u$
&\multicolumn{1}{c}{r}&\multicolumn{1}{c}{$x$}&$\sigma_{x}$ &\multicolumn{1}{c}{$y$}&$\sigma_{y}$&P&S&Match ID\\
 & \multicolumn{2}{c}{(km s$^{-1}$)}& (mas) &\multicolumn{2}{c}{(mas)}&\multicolumn{2}{c}{(mas)}&(Jy)&(Jy km s$^{-1}$)&Epoch 2\\
(1)&(2)&(3)&\multicolumn{1}{c}{(4)}&(5)&\multicolumn{1}{c}{(6)}&(7)&(8)&(9)&(10)&(11)
}\startdata

      1&           -17.28&       1.52& 8.10 &     -2.862&     0.002&    -0.452&     0.005&        11.7&      36.8&      1\\
      2$^{\ast}$&  -16.32&       0.65& 6.96 &     -1.356&     0.009&     7.974&     0.032&       0.7&       2.1&  ... \\
      3&           -15.16&       3.26& 8.31 &     -1.637&     0.004&     0.448&     0.008&           19.7&     147.5&      2\\
      4$^{\ast}$&  -14.94&       0.87& 8.76 &     -1.679&     0.007&    -0.250&     0.022&           1.5&       4.9&  ...    \\
      5&           -14.54&       2.82& 8.42 &     -1.304&     0.003&     0.686&     0.008&          16.6&      84.3&      6\\
      6$^{\ast}$&  -14.16&       0.87& 8.56 &     -1.724&     0.004&    -0.018&     0.015&        8.2&      21.0&  ...    \\
      7&           -13.23&       0.87& 8.34 &     -3.051&     0.004&    -0.909&     0.009&         3.1&       8.1&      8\\
      8&           -12.56&       1.09& 8.45 &     -1.228&     0.007&     0.738&     0.007&        10.3&      28.7&     11\\
      9&           -12.22&       3.04& 8.87 &     -0.792&     0.003&     0.621&     0.005&        25.5&     151.3&      9\\
     10&           -11.98&       2.17& 8.38 &     -1.790&     0.002&     0.176&     0.007&         9.4&      52.9&      ...\\
     11&           -10.94&       2.17& 9.47 &     -0.330&     0.011&     0.243&     0.015&         9.8&      59.9&     13\\
     12&            -9.97&       1.30& 9.04 &     -1.329&     0.008&    -0.305&     0.019&         4.5&      18.9&     14\\
     13&            -9.77&       2.82& 9.30 &     -0.957&     0.003&    -0.280&     0.008&        15.9&     103.8&     16\\
     14&            -9.57&       1.09& 8.39 &     -16.011&     0.007&     8.277&     0.013&         5.2&      21.6&     ...\\
     15$^{R}$&      -9.31&       2.39& 9.88 &     0.000&     0.005&     0.000&  0.007&        35.9&     254.8&     18\\
     16&            -9.20&       1.09& 9.58 &     0.030&     0.019&     0.567&     0.012&         1.9&       6.6&     19\\
     17$^{\ast}$&   -9.05&       1.09& 8.95 &     -15.261&     0.009&      11.052&     0.036&     3.5&      10.1&     20\\
     18$^{\ast}$&   -9.04&       0.87& 8.93 &     -15.099&     0.034&    11.227&     0.021&       5.1&      12.7&  ...    \\
     19&            -8.96&       1.95& 9.39 &     -15.667&     0.006&    11.237&     0.018&           6.2&      29.5&     21\\
     20&            -8.68&       1.74& 10.21&     0.412&     0.008&     0.014&     0.026&           6.7&      23.3&     22\\
     21&            -8.58&       1.09& 11.93&     -16.001&     0.009&    14.662&     0.029&           1.4&       6.0&     23\\
     22&            -8.33&       2.82& 8.37 &     -15.791&     0.009&     8.848&     0.016&           5.3&      40.4&     27\\
     23$^{\ast}$&   -8.32&       0.65& 11.59&     -1.897&     0.012&    -4.038&     0.028&            1.2&       3.1&  ...    \\
     24&            -8.23&       1.52& 9.79 &     -12.858&     0.006&    14.322&     0.019&            4.4&      19.2&     25\\
     25&            -8.12&       1.09& 8.58 &     -14.846&     0.005&    10.985&     0.014&            7.5&      26.2&     24\\
     26&            -8.02&       2.82& 9.74 &      0.438&     0.009&     0.925&     0.018&            7.2&      45.9&     26\\
     27&            -7.80&       1.74& 7.91 &     -9.327&     0.007&    13.627&     0.013&            5.3&      29.1&     29\\
     28&            -7.56&       1.09& 11.10&     -15.678&     0.024&    13.829&     0.041&            1.5&       6.3& ...     \\
     29&            -7.55&       1.09& 7.13 &     -7.390&     0.009&    12.937&     0.029&            1.9&       6.9&  ...    \\
     30&            -7.07&       1.74& 11.44&     -15.386&     0.006&    14.560&     0.015&            4.3&      22.1&     30\\
     31&            -6.72&       1.09& 8.24 &     -9.195&     0.010&    13.984&     0.025&            2.3&       8.6&     31\\
     32&            -6.70&       2.39& 6.90 &     -5.873&     0.003&    12.406&     0.014&            8.8&      53.1&  ...    \\
     33&            -6.45&       1.74& 9.34 &     -12.505&     0.012&    14.010&     0.018&            2.8&      14.8&     ...\\
     34$^{\ast}$&   -6.14&       0.65& 7.81 &     -7.069&     0.011&    13.593&     0.032&           1.2&       2.8&  ...    \\
     35$^{\ast}$&   -6.10&       0.65& 9.38 &     -16.537&     0.015&     9.687&     0.032&           1.8&       4.9&     32\\
     36&            -5.53&       1.52& 8.82 &     -8.159&     0.005&    14.657&     0.003&         22.6&      92.2&     34\\
     37&            -4.58&       1.74& 9.07 &     -16.715&     0.003&     8.303&     0.006&         10.3&      48.5&     36\\
     38&            -4.20&       1.09& 8.88 &     -7.921&     0.006&    14.718&     0.003&          8.6&      41.4&     37\\
     39$^{\ast}$&   -3.70&       0.65& 9.51 &     -17.329&     0.024&     4.082&     0.039&           2.4&       6.4&     39\\
     40&            -3.10&       2.39& 8.75 &     -16.673&     0.007&     4.814&     0.016&           11.2&      67.2&     43\\
     41&            -3.03&       1.96& 9.13 &     -7.780&     0.006&    14.967&     0.008&           12.4&      75.7&     41\\
     42&            -2.88&       2.17& 9.25 &     -16.938&     0.008&     8.141&     0.009&            6.7&      44.6&     42\\
     43&            -2.82&       1.74& 8.39 &     -16.237&     0.013&     4.292&     0.024&            3.5&      13.5&     47\\
     44&            -2.78&       1.52& 11.08&    -14.008&     0.016&    15.137&     0.026&            2.2&      10.0&     46\\
     45&            -2.32&       2.39& 9.17 &     -16.406&     0.006&     9.461&     0.009&           12.9&      79.8&     45\\
     46&            -1.72&       2.17& 8.03 &     -15.938&     0.009&     4.691&     0.018&            7.2&      46.0&     48\\
     47$^{\ast}$&   -1.50&      0.87&  8.27 &     -16.227&     0.022&     5.114&     0.047&           4.6&       6.4&     50\\
     48$^{\ast}$&   -0.92&      0.65&  7.73 &     -15.366&     0.015&     3.546&     0.043&           2.0&       5.5&  ...    \\
     49&            -0.73&       1.52& 8.66 &     -15.946&     0.017&     9.235&     0.038&           3.8&      15.5&     ...\\
     50&            -0.64&       1.52& 9.15 &     -11.534&     0.012&    -2.610&     0.022&           4.2&      18.8&     52\\
     51&            -0.57&       1.09& 9.96 &      1.832&     0.021&     7.554&     0.029&           1.6&       7.2&     53\\
     52&            0.25&       1.09&  7.10 &     -14.770&     0.015&     3.747&     0.032&           2.1&       7.6&     54\\
     53&            1.08&       2.17&  8.67 &     -11.042&     0.014&    -2.283&     0.025&           5.4&      30.3&     57\\
     54$^{\ast}$&   1.30&       0.65&  8.07 &     -11.478&     0.047&    13.113&     0.047&        4.8&       8.2&  ...    \\
     55&            1.43&       1.30&  10.73 &      2.032&     0.010&     1.950&     0.014&        3.5&      14.1&     56\\
     56&            1.44&        1.09& 8.50 &     -10.621&     0.005&    -2.252&     0.016&        3.5&      10.1&     55\\
     57$^{\ast}$&   1.48&       0.65&  8.23 &     -11.841&     0.012&    13.107&     0.050&            1.8&       4.3&     58\\
     58&            2.34&       1.52&  7.57 &     -11.300&     0.005&    12.637&     0.021&         8.4&      36.4&     61\\
     59&            2.65&       1.95&  8.11 &     -12.171&     0.009&    12.774&     0.030&         3.1&      20.3&   ...   \\
     60&            2.87&       1.74&  7.72 &     -2.150&     0.014&     0.760&     0.021&         1.8&      11.9&     65\\
     61&            3.19&       1.09&  7.08 &     -11.118&     0.018&    12.181&     0.023&         7.4&      24.4&     64\\
     62&            3.77&       1.52&  6.84 &     -10.486&     0.006&    12.202&     0.016&         4.0&      16.7&     66\\
     63&            3.86&       1.52&  7.73 &     -12.410&     0.021&    12.175&     0.024&         3.1&      17.1&     ...\\
     64$^{\ast}$&   3.86&       0.65&  7.38 &     -10.185&     0.010&    -1.221&     0.045&           1.6&       4.0&     68\\
     65&            4.04&       1.30&  7.43 &     -1.074&     0.007&     3.081&     0.023&          1.2&       6.0&   ...   \\
     66&            4.10&       3.04&  8.66 &      0.508&     0.008&     4.077&     0.013&          2.8&      29.4&     70\\
     67&            4.70&       1.52&  7.40 &     -12.565&     0.009&    11.650&     0.020&          5.8&      30.0&     71\\
     68&            4.83&       1.30&  6.67 &     -9.377&     0.012&    -0.696&     0.033&          2.3&      10.2&   ... \\
     69&            5.20&       1.95&  7.22 &     -3.566&     0.014&     0.108&     0.027&          2.4&      14.6&   ...\\
     70&            5.43&       1.30&  6.05 &     -10.355&     0.003&    11.399&     0.012&          5.5&      19.7&     75\\
     71$^{\ast}$&   5.59&        0.65& 6.82 &     -4.651&     0.008&    -0.124&     0.023&           1.2&       3.2&     76\\
     72&            5.97&       1.52&  5.75 &     -13.283&     0.009&     3.593&     0.016&           2.5&      15.5&     74\\
     73&            7.00&       1.30&  6.21 &     -7.430&     0.011&    12.023&     0.025&           1.9&       8.1&     79\\
     74&            7.33&       3.04&  6.39 &     -9.017&     0.006&    -0.477&     0.013&           4.0&      41.0&     78\\
     75&            7.42&       1.74&  6.51 &     -8.278&     0.008&    12.338&     0.022&           2.9&      16.6&   ...\\
     76&            7.73&       2.17&  5.08 &     -12.707&     0.005&     3.946&     0.008&           4.0&      27.5&     82\\
     77$^{\ast}$&   7.99&       0.65&  7.79 &     -0.359&     0.017&     4.208&     0.034&           1.3&       3.5& ... \\
     78$^{\ast}$&   8.01&       1.09&  6.16 &     -8.120&     0.008&    11.995&     0.023&           2.3&       7.8&     83\\
     79$^{\ast}$&   8.35&       1.09&  3.98 &     -4.070&     0.014&     6.607&     0.026&           1.5&       6.5& ... \\
     80&            8.92&       1.95&  5.85 &     -8.965&     0.006&     0.059&     0.023&          4.9&      31.9&     84\\
     81&            9.17&       1.30&  5.85 &     -8.289&     0.007&    11.678&     0.013&          3.9&      17.9&     85\\
     82&           12.62&       0.87&  2.50 &     -5.479&     0.009&     5.706&     0.017&          2.2&       7.4&     86\\

\enddata
\tablecomments{column (1): ID number; columns (2): V$_{LSR}$ at
the peak of velocity profile of feature; column (3): the velocity
range across the feature $\Delta u$; column (4): distance of maser
feature, $r$, from the fitted position of central star; columns
(5) and (7): the intensity weighted centroid of each feature ($x,
y$); columns (6) and (8): the corresponding uncertainties
($\sigma_{x}, \sigma_{y}$); column (9): the peak flux density of
each feature $P$; column (10): the integrated flux density of all
spots in the feature S; and column (11): the ID numbers of matched
features at another epoch.}

\tablenotetext{*}{Feature which can not be well
represented by a Gaussian curve.}

\tablenotetext{R} {Reference feature.}
\end{deluxetable}

\begin{deluxetable}{lrrrrrrrrrrrrr}
\tabletypesize{\tiny} \setlength{\tabcolsep}{0.11in}

\tablewidth{0pt} \tablecaption{\footnotesize 43 GHz SiO maser
features around AH Sco observed by VLBA on March 20, 2004.}

\tablehead{ID&\multicolumn{1}{r}{V$_{LSR}$}&$\Delta u$
&\multicolumn{1}{c}{r}&\multicolumn{1}{c}{$x$}&$\sigma_{x}$ &\multicolumn{1}{c}{$y$}&$\sigma_{y}$&P&S&Match ID\\
 & \multicolumn{2}{c}{(km s$^{-1}$)}& (mas) &\multicolumn{2}{c}{(mas)}&\multicolumn{2}{c}{(mas)}&(Jy)&(Jy km s$^{-1}$)&Epoch 2\\
(1)&(2)&(3)&\multicolumn{1}{c}{(4)}&(5)&\multicolumn{1}{c}{(6)}&(7)&(8)&(9)&(10)&(11)
}\startdata

      1&           -17.23&      1.09& 8.14 &  -2.861&     0.003&    -0.509&     0.014&          5.8&      16.0&      1\\
      2&           -15.14&      2.39& 8.33 &  -1.650&     0.004&     0.385&     0.009&           14.1&      90.0&      3\\
      3$^{\ast}$&  -14.73&      0.87& 7.97 &  -1.642&     0.012&     0.953&     0.021&           2.5&       3.8&  ...    \\
      4$^{\ast}$&  -14.60&      0.87& 8.03 &  -1.351&     0.018&     1.243&     0.011&           1.3&       3.4&  ...    \\
      5$^{\ast}$&  -14.59&      0.65& 8.90 &  -1.271&     0.010&    -0.055&     0.011&           1.2&       3.1&  ...    \\
      6&           -14.50&      2.39& 8.39 &  -1.351&     0.003&     0.642&     0.009&       21.7&     121.4&      5\\
      7$^{\ast}$&  -13.74&      0.65& 9.04 &  -0.905&     0.052&     0.159&     0.091&           2.1&       5.5&  ...    \\
      8$^{\ast}$&  -13.31&      0.65& 8.37 &  -3.046&     0.005&    -0.946&     0.017&           2.0&       4.4&      7\\
      9&           -12.41&      4.13& 8.85 &  -0.813&     0.006&     0.591&     0.015&           18.9&     144.1&      9\\
     10&           -12.17&      2.82& 8.44 &  -1.777&     0.004&     0.080&     0.016&            6.6&      42.7&      ...\\
     11&           -12.14&      1.52& 8.56 &  -1.139&     0.016&     0.640&     0.012&           15.4&      61.6&      8\\
     12&           -11.07&      0.87& 9.09 &  -0.930&     0.010&     0.054&     0.018&            7.9&      24.6&   ... \\
     13&           -10.89&      1.74& 9.33 &  -0.423&     0.021&     0.318&     0.014&           11.1&      53.6&     11\\
     14&           -10.08&      1.52& 8.99 &  -1.366&     0.006&    -0.296&     0.022&            5.8&      20.6&     12\\
     15$^{\ast}$&  -9.88&       0.87& 9.80 &  -0.906&     0.017&    -0.988&     0.021&           1.4&       3.6&  ... \\
     16&           -9.72&       3.48& 9.27 &  -1.002&     0.005&    -0.301&     0.010&          14.7&     109.9&     13\\
     17&           -9.50&       1.30& 8.36 &  -15.994&     0.010&     8.099&     0.029&           3.6&      18.3&     ...\\
     18$^{R}$&     -9.36&       3.04& 9.86 &  0.000&     0.005&     0.000&     0.008&           32.0&     218.4&     15\\
     19&           -9.16&       1.52& 9.54 &  -0.015&     0.007&     0.519&     0.017&       1.9&       7.4&     16\\
     20&           -8.98&       1.30& 8.97 &   -15.241&     0.007&    11.070&     0.021&       5.5&      21.6&     17\\
     21&           -8.83&       2.61& 9.35 &   -15.673&     0.007&    11.118&     0.029&       7.5&      38.1&     19\\
     22$^{\ast}$&  -8.55&       0.65& 10.21&   0.354&     0.006&    -0.109&     0.026&         5.9&      11.4&     20\\
     23&           -8.11&       1.09& 11.85&   -16.001&     0.006&    14.542&     0.021&          2.6&      11.5&     21\\
     24&           -8.09&       1.52& 8.60 &  -14.867&     0.006&    10.958&     0.016&          8.8&      36.0&     25\\
     25&           -8.08&       1.52& 9.75 &  -12.873&     0.009&    14.265&     0.026&          3.1&      15.2&     24\\
     26&           -8.07&       1.96& 9.72 &    0.428&     0.012&     0.887&     0.027&          6.7&      37.0&     26\\
     27&           -8.00&       2.17& 8.39 &  -15.780&     0.009&     8.829&     0.017&          9.6&      49.4&     22\\
     28&           -7.75&       1.96& 6.78 &   -5.908&     0.005&    12.333&     0.017&          4.2&      22.0&   ...\\
     29&           -7.41&       2.17& 7.81 &   -9.290&     0.009&    13.558&     0.024&          3.8&      21.6&     27\\
     30&           -7.09&       1.30& 11.35&  -15.398&     0.011&    14.414&     0.031&          3.2&      17.1&     30\\
     31$^{\ast}$&  -6.99&       1.30& 8.21 &  -9.175&     0.012&    13.978&     0.033&      3.2&       7.1&     31\\
     32&           -6.34&       1.52& 9.39 &  -16.524&     0.017&     9.659&     0.045&      2.2&      11.9&     35\\
     33&           -6.31&       1.52& 9.17 &  -12.445&     0.012&    13.846&     0.027&      2.4&      12.8&     ...\\
     34&           -5.74&       1.30& 8.74 &  -8.193&     0.002&    14.598&     0.007&     18.7&      55.1&     36\\
     35$^{\ast}$&  -5.51&       0.65& 8.24 &  -8.555&     0.022&    14.083&     0.071&       1.8&       4.8&   ...\\
     36&           -4.62&       1.74& 9.12 &  -16.734&     0.004&     8.241&     0.011&       6.2&      33.0&     37\\
     37&           -4.30&       1.74& 8.76 &  -7.981&     0.010&    14.622&     0.009&      12.4&      69.8&     38\\
     38&           -4.08&       1.52& 9.47 &  -7.822&     0.014&    -3.618&     0.048&       2.1&      11.2&   ...\\
     39$^{\ast}$&  -3.72&       0.65& 9.58 &  -17.347&     0.037&     4.073&     0.048&       2.8&       7.5&     39\\
     40$^{\ast}$&  -3.30&       0.65& 10.93&  -13.892&     0.019&    15.022&     0.059&       1.2&       3.0&  ... \\
     41&           -3.16&       2.61& 9.03 &  -7.811&     0.009&    14.892&     0.019&       9.4&      62.0&     41\\
     42&           -3.09&       2.39& 9.26 &  -16.914&     0.011&     8.125&     0.023&       3.9&      26.3&     42\\
     43&           -3.05&       2.17& 8.75 &  -16.624&     0.011&     4.791&     0.021&       9.0&      60.1&     40\\
     44$^{\ast}$&  -2.67&       0.65& 8.33 &  -7.722&     0.033&    14.190&     0.027&      1.2&       2.9& ...  \\
     45&           -2.40&       2.61& 9.14 &  -16.370&     0.008&     9.381&     0.023&       13.2&      95.3&     45\\
     46$^{\ast}$&  -2.40&       0.87& 11.11&  -14.067&     0.013&    15.123&     0.049&      1.4&       3.6&     44\\
     47$^{\ast}$&  -2.36&       1.09& 8.42 &  -16.199&     0.033&     4.248&     0.097&      4.6&      14.1&     43\\
     48&           -1.83&       1.96& 8.08 &  -15.931&     0.013&     4.670&     0.034&      5.2&      38.6&     46\\
     49$^{\ast}$&  -1.61&       0.65& 8.63 &   -16.528&     0.012&     5.067&     0.036&      2.2&       6.0&  ... \\
     50$^{\ast}$&  -1.16&       0.65& 8.27 &   -16.157&     0.011&     5.014&     0.022&      2.0&       5.1&     47\\
     51&           -0.54&       1.30& 8.48 &  -15.780&     0.019&     9.080&     0.040&      1.7&      12.6&     ...\\
     52&           -0.47&       1.74& 9.27 &  -11.556&     0.008&    -2.679&     0.026&      3.0&      18.7&     50\\
     53&           -0.33&       0.87& 9.83 &  1.767&     0.033&     7.487&     0.068&      2.2&       6.2&     51\\
     54$^{\ast}$&   0.39&       0.65& 7.12 &  -14.744&     0.034&     3.762&     0.065&       1.8&       4.1&     52\\
     55&            1.49&       1.30& 8.52 &  -10.627&     0.009&    -2.226&     0.030&      3.6&      14.6&     56\\
     56&            1.49&       1.30& 10.65&   1.969&     0.011&     1.927&     0.019&      2.7&      12.1&     55\\
     57&            1.55&       1.96& 8.71 &  -11.052&     0.013&    -2.276&     0.033&      3.4&      24.7&     53\\
     58&            1.58&       1.30& 8.10 &  -11.766&     0.013&    12.997&     0.035&      3.7&      13.7&     57\\
     59$^{\ast}$&   1.70&       0.65& 9.30 &  -1.642&     0.023&    12.725&     0.052&     1.3&       3.1&  ... \\
     60&            1.85&       0.87& 8.33 &   -12.225&     0.033&    13.001&     0.036&    5.3&       9.5&   ...\\
     61&            2.30&       1.95& 7.55 &   -11.316&     0.010&    12.614&     0.023&    8.0&      52.1&     58\\
     62&             2.85&      1.74& 8.01 &   -12.161&     0.008&    12.666&     0.029&    2.5&      13.7&   ...\\
     63&             3.07&      1.09& 8.78 &    0.646&     0.006&     3.962&     0.020&    2.3&       8.7&  ...\\
     64&             3.19&      1.74& 7.03 &   -11.131&     0.017&    12.125&     0.029&   10.8&      34.3&     61\\
     65&             3.35&      1.30& 7.68 &   -2.163&     0.020&     0.787&     0.047&    1.3&       6.8&     60\\
     66&             3.50&      1.52& 6.85 &  -10.396&     0.011&    12.258&     0.025&    5.1&      23.6&     62\\
     67&            4.13&       1.95& 7.66 &  -12.283&     0.017&    12.167&     0.028&    2.9&      12.7&     ...\\
     68&            4.17&       0.87& 7.46 &  -10.183&     0.011&    -1.255&     0.053&    1.7&       5.5&     64\\
     69&            4.27&       1.09& 7.42 &   -3.501&     0.016&    -0.097&     0.053&    1.4&       5.1&   ... \\
     70&            4.29&       1.74& 8.55 &    0.434&     0.010&     4.042&     0.026&    2.5&      14.7&     66\\
     71&            4.78&       1.52& 7.37 &  -12.552&     0.013&    11.604&     0.020&    8.0&      35.3&     67\\
     72&            5.04&       1.74& 6.40 &   -1.921&     0.007&     8.073&     0.026&    2.5&      13.7&  ... \\
     73$^{\ast}$&   5.44&       0.87& 6.93 &  -3.062&     0.011&     0.916&     0.038&    1.2&       3.6&  ... \\
     74&            5.56&       1.30& 5.79 &  -13.234&     0.012&     3.534&     0.023&     2.9&      12.2&     72\\
     75&            5.67&       1.52& 5.94 &  -10.333&     0.004&    11.302&     0.014&     5.5&      22.0&     70\\
     76$^{\ast}$&   5.72&       0.87& 6.84 &   -4.629&     0.011&    -0.132&     0.038&       1.5&       5.7&     71\\
     77&            6.06&       0.87& 7.13 &   -3.665&     0.008&     0.137&     0.026&     1.7&       5.4&    ...\\
     78&            7.05&       1.74& 6.46 &   -8.987&     0.013&    -0.517&     0.027&     5.3&      25.7&     74\\
     79&            7.23&       1.30& 6.08 &   -7.467&     0.014&    11.923&     0.042&     1.4&       6.3&     73\\
     80&            7.51&       1.09& 7.87 &   -0.253&     0.016&     4.125&     0.048&     1.9&       6.6&  ... \\
     81&            7.91&       1.74& 6.48 &   -9.070&     0.010&    -0.522&     0.033&     4.7&      26.1&   ... \\
     82&            7.92&       2.39& 5.13 &  -12.679&     0.008&     3.906&     0.022&     2.8&      19.6&     76\\
     83&            8.09&       2.39& 6.21 &   -8.201&     0.013&    12.065&     0.036&     3.7&      25.2&     78\\
     84&            9.15&       1.96& 5.89 &   -8.943&     0.006&     0.052&     0.020&     5.1&      31.3&     80\\
     85&            9.31&       1.30& 5.73 &   -8.309&     0.008&    11.587&     0.017&     3.5&      16.1&     81\\
     86&           12.98&       1.09& 2.41 &   -5.522&     0.008&     5.632&     0.026&     2.9&      10.8&     82\\
     87&           12.99&       1.09& 4.44 &   -6.258&     0.007&     9.977&     0.029&     1.8&       6.4&   ...\\

\enddata

\tablecomments{The representations of columns (1)-(11) are the
same as in Table 1.}
 \tablenotetext{*}{Feature which can not be well
represented by a Gaussian curve.}

\tablenotetext{R} {Reference feature.}

\end{deluxetable}

\clearpage


\begin{deluxetable}{lccl}
\tabletypesize{\normalsize}
 \setlength{\tabcolsep}{0.10in}

\tablewidth{0pt} \tablecaption{Sources with the detected
contraction of SiO maser shell.}

\tablehead{Source & stellar phase & Contraction velocity &
Reference\\
 & $\phi$ & km s$^{-1}$ &}
\startdata

R Aqr & $0.78-1.04$ & 4.2$\pm$0.9 & Boboltz et al. (1997)\\
TX Cam & $0.50-0.65$ & $5-10$ & Diamond \& Kemball (2003)\\
VX Sgr & $0.75-0.80$ & 4.1$\pm$0.6 & Chen et al. (2006)\\
AH Sco & $\sim$ 0.55 & 12.7$\pm$0.7 & this work \\

\enddata
\end{deluxetable}


\begin{deluxetable}{ccccccc}
\tabletypesize{\small} \setlength{\tabcolsep}{0.06in}
\tablewidth{0pt} \tablecaption{Best-fit model for the SiO maser
kinematics in AH Sco.}
\tablehead{$v_{0x}$ & $v_{0y}$ & $v_{0z}$ & $V_1$ & $\alpha$ & d & Reduced $\chi^2$\\
(km s$^{-1}$)&(km s$^{-1}$)&(km s$^{-1}$) &(km s$^{-1}$)& &(kpc)&}

\startdata
$3.8\pm0.3$ & $-0.9\pm0.4$ & $-5.2\pm0.4$ & $-14.1\pm1.4$ & $-0.54\pm0.16$ & $2.26\pm0.19$& 3.58 \\

\enddata

\end{deluxetable}

\clearpage

\begin{figure*}
\begin{center}
\includegraphics[scale=1.2]{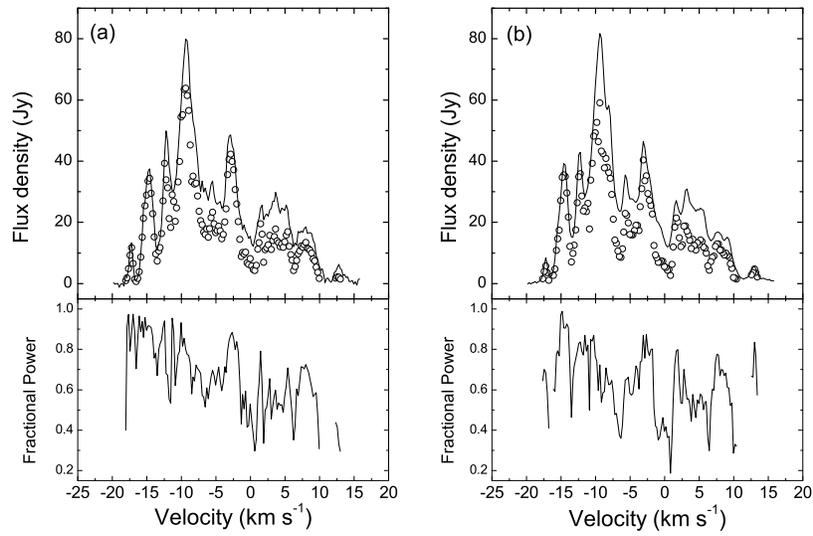}

\caption{Top: Comparison of total power (solid line) to cross
power (open circle) of 43 GHz {\it v}=1, {\it J}=1--0 SiO maser
emission toward AH Sco obtained on (a) March 8, 2004 and (b) March
20, 2004. Bottom: The corresponding fraction power (cross/total)
detected by the high-resolution VLBA observations. }
\end{center}
\end{figure*}

\begin{figure*}
\includegraphics[scale=0.9]{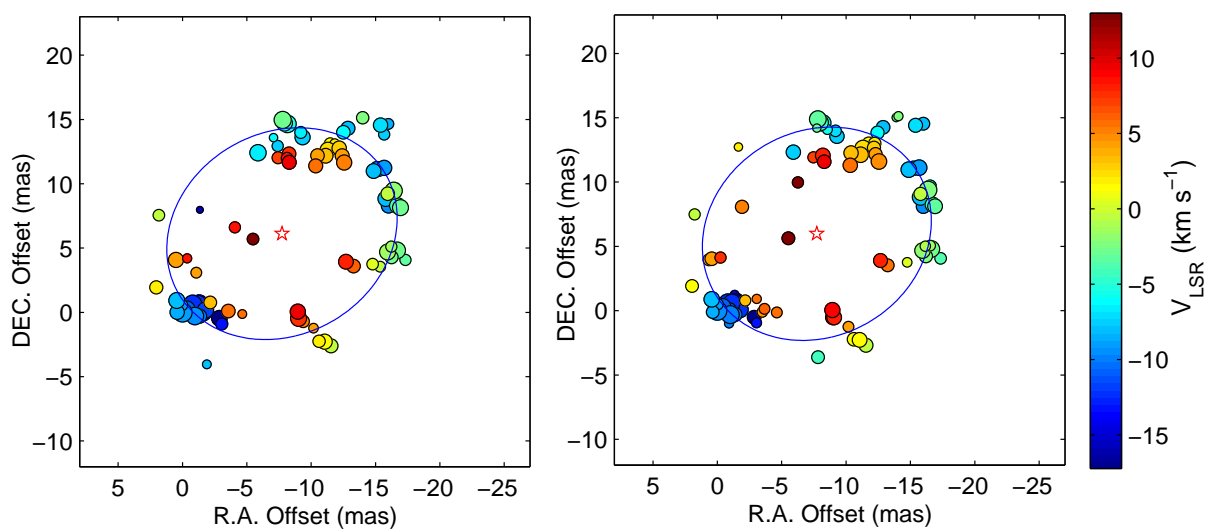}

\caption{VLBI images of 43 GHz {\it v}=1, {\it J}=1--0 SiO maser
emission toward AH Sco obtained on (a) March 8, 2004 and (b) March
20, 2004. Each maser feature is represented by a filled circle
whose area is proportional to the logarithm of the flux density,
and the color indicates its Doppler velocity with respect to the
local standard of rest. Its stellar velocity is about $-7$ km
s$^{-1}$. Errors in the positions of the features are smaller than
the data points. The ellipse indicates the least-squares fit to
the maser distribution for each epoch. The fitted center of
ellipse model is marked by the red star.}

\end{figure*}

\begin{figure*}
\begin{center}
\includegraphics[scale=0.9]{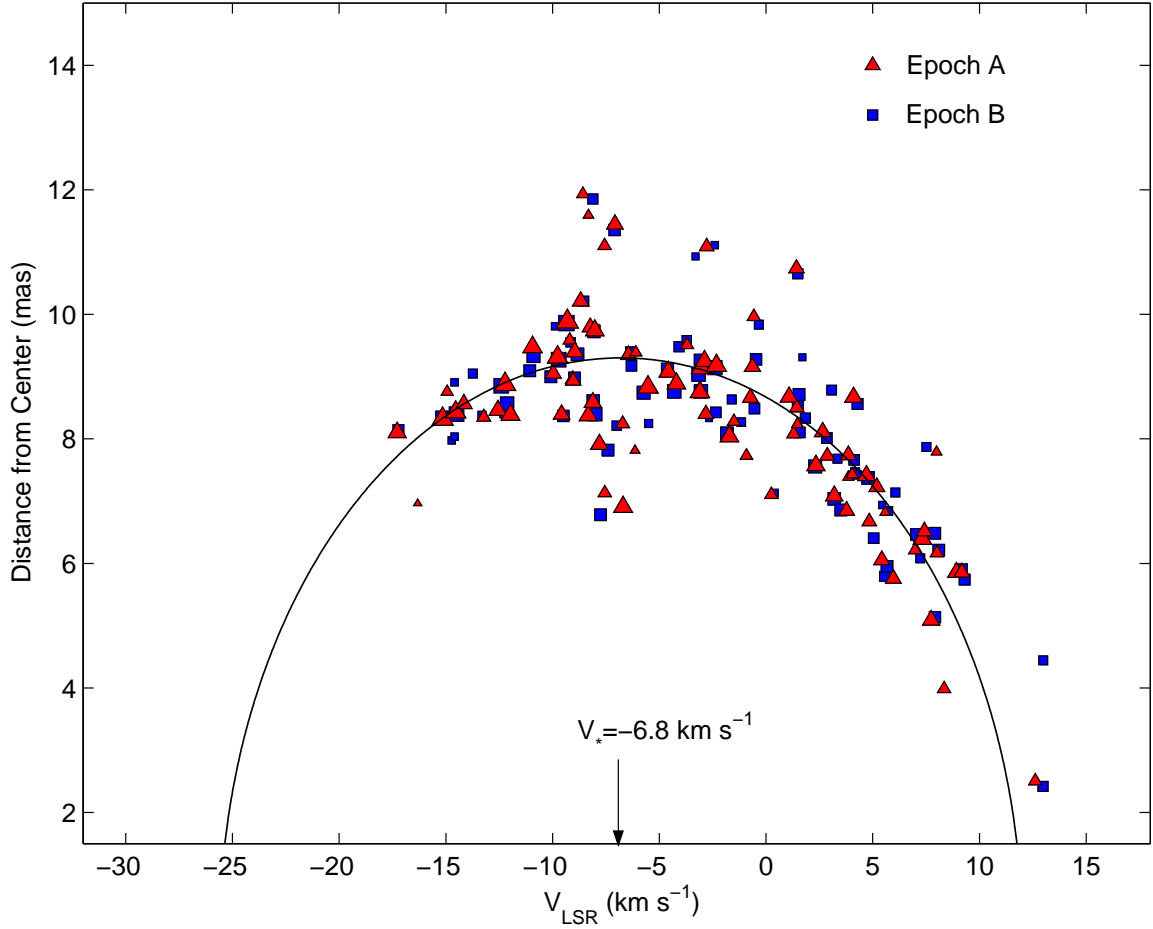}

\caption{Distance of maser features from the fitted position of
central star (in Fig. 2) versus their LOS velocity for epochs A
and B. Maser features of the different epochs are denoted by
different color symbols whose area is proportional to the
logarithm of the flux density. The plot suggests that the
higher-velocity maser features lie closer to the central star,
which can be well explained by the uniformly expanding thin shell
model. Indicated by the downward arrow is the systemic velocity of
AH Sco of $-6.8$ km s$^{-1}$, obtained from the best-fitting
thin-shell model (shown by the curve).}
\end{center}
\end{figure*}

\begin{figure*}
\begin{center}
\includegraphics[scale=1]{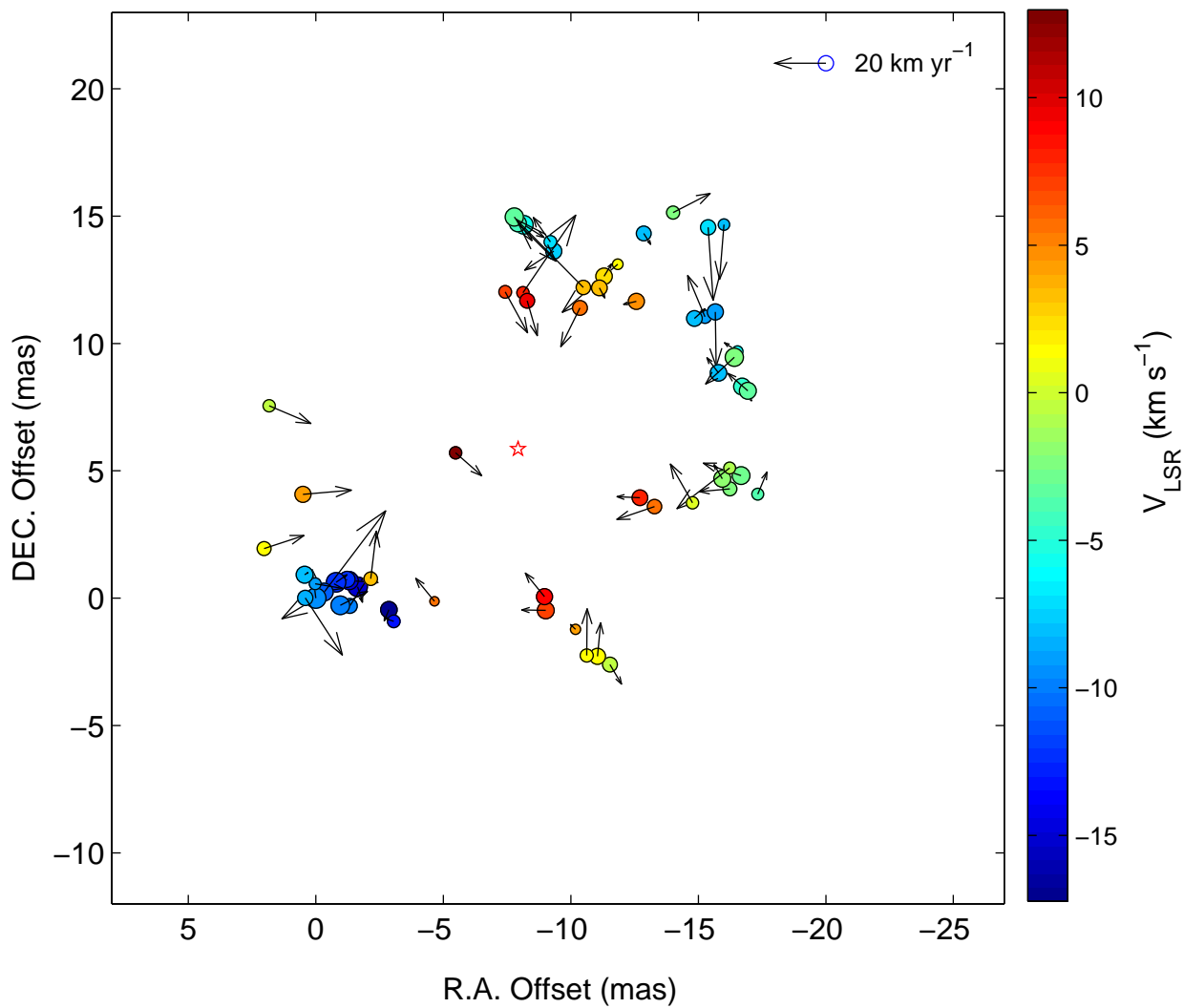}

\caption{Distribution of proper motion velocity vectors of the
matched maser features at an assumed distance of 2.26 kpc. The
length of the vector is proportional to the velocity. The mean
proper motion vector has been subtracted from each of the
determined proper motion vectors. The color and size of symbols
are the same as that shown in Fig. 2. Red star represents the
fitted center of ellipse model to maser distribution (see Sect.
3.1).}
\end{center}
\end{figure*}

\end{document}